\documentclass[%
 reprint,
amsmath,amssymb,
aps,
]{revtex4-2}

\usepackage{graphicx}
\usepackage{dcolumn}
\usepackage{bm}
\usepackage{orcidlink} 
\usepackage{float}

\begin{document}

\title{\boldmath Neutrino oscillations with atmospheric neutrinos at large liquid argon TPCs }

\author{Animesh~Chatterjee$^{1,2}$ \orcidlink{0000-0002-2935-0958}}
\email{animesh.chatterjee@cern.ch }
\author{Albert De Roeck$^1$ \orcidlink{0000-0002-5614-4092}}
\email{Albert.de.Roeck@cern.ch}

\affiliation{$^1$ European Organization for Nuclear Research (CERN), 1211 Geneva 23, Switzerland}
\affiliation{$^2$ Physical Research Laboratory, Ahmedabad, Gujarat, 380009, India}

\begin{abstract}
    We propose to study atmospheric neutrino interactions with a unique event topology to distinguish neutrinos and anti-neutrinos using a liquid argon time projection chamber in an experiment such as DUNE. The detection of CC1P and CC0P events will allow to access neutrino oscillation physics complementary to accelerator based beam neutrinos. Our analysis shows that a sensitivity to the mass-ordering can be achieved with a significance close to 4$\sigma$ and a CP violation sensitivity with more than 2$\sigma$ with a data sample of 140 kt-yr of atmospheric neutrinos in the DUNE detector.
\end{abstract}

\maketitle
\flushbottom

\section{Introduction}
\label{sec:intro}
    Neutrino oscillations, in which one flavour of neutrinos  converts into others, have been discovered using a variety of neutrino sources: from neutrinos produced in the sun to atmospheric neutrinos, from reactors to accelerators, and with a variety of detection techniques in different terrestrial experiments\cite{McDonald:2016ixn,Kajita:2016cak}.\\ 
    In 1998 data from atmospheric neutrinos at the Super-Kamiokande (SK) experiment clearly established neutrino oscillations\cite{Super-Kamiokande:2004orf}. This also established the phenomenon of neutrino masses and mixing, and thus provided the first evidence of physics beyond the standard model.\\
    A major activity in neutrino oscillation experiments over the past two decades has focused on the precise determination of a) the neutrino mass-squared differences, b)  mixing angles and c) extracting the CP phase $\delta_{CP}$\cite{Pontecorvo:1967fh,Gribov:1968kq}.The future long-baseline facility Deep Underground Neutrino Experiment (DUNE)\cite{DUNE:2020ypp,DUNE:2020txw} aims to extract the sign of the atmospheric mass splitting and the CP-violating phase $\delta_{cp}$ through the golden channels $\nu_{\mu}\rightarrow \nu_{e}$ and $\bar{\nu_{\mu}}\rightarrow \bar{\nu_{e}}$. However, both quantities can also be measured using atmospheric neutrinos as proposed in the literature since a long time\cite{Banuls:2001zn,Gandhi:2007td,Gonzalez-Garcia:2004pka,Choubey:2006jk,Arguelles:2022hrt,Kelly:2019itm}.\\
    Although high energy atmospheric neutrinos were originally the focus of interest, the wide energy range ($\mathcal{O}(10^{-2})$ to $\mathcal{O}(10^{5})$ GeV) and large distances ($L$ of the range from 17 up to almost 12800 km) covered, makes it  one of the golden samples to study neutrino oscillations as well as other beyond the standard model physics.\\ 
     In this letter, we are particularly interested to explore the possibility of using atmospheric neutrino events from sub-GeV ($< $1 GeV) to 10s of GeV, using a liquid argon time projection chambers (LArTPCs). The primary reasons are the following:
     a) Low energy (sub-GeV) atmospheric neutrinos, which cover a large distance (earth radius) L, will have a significant effect on the oscillation probability arising from both the atmospheric as well as from the solar mass-splitting. It has been shown earlier\cite{Kelly:2019itm, Indumathi:2017kxa} that the impact on the oscillation probabilities of the leptonic CP violating phase $\delta_{CP}$ are much  stronger in sub-GeV atmospheric neutrinos as compared to long-baseline accelerator-based beam neutrinos. Atmospheric neutrino experiments collect both $\nu$ and $\bar{\nu}$ and a measurement of the oscillation pattern of  the neutrinos and anti-neutrinos with the wide energy range (from sub-GeV to 10s of GeV) can provide important new information on the measurement of $\delta_{CP}$ if the 
     incoming neutrino can be tagged. b) The combination of large $\theta_{13}$, separation of neutrino and anti-neutrinos till a few GeV  (as discussed below),the wide neutrino energy range and large matter effects will help significantly to resolve mass ordering\cite{Gandhi:2008zs,Barger:2012fx}. c) the pioneering technique of liquid argon time projection chambers (LArTPCs)\cite{Rubbia:1977zz} 
     with its full 3D-imaging, excellent particle identification (PID) capability, and precise calorimetric energy reconstruction will play a key role in identifying neutrino event topologies. This unique technique will help to separate neutrino and anti-neutrino samples coming from atmospheric neutrinos, as discussed below. Therefore, such measurements will not only have a significant impact on the CP violation but also on resolving the mass ordering and octant degeneracy.\\ 
    Inspired by the previous study on the CP violation using sub-GeV atmospheric neutrinos at DUNE\cite{Kelly:2019itm}, in this Letter, we propose to study the atmospheric neutrinos (with a wide range of energies from 100 MeV  to 10 GeV), exploiting the unique event topology to distinguish neutrinos and anti-neutrinos using the liquid argon time projection chamber at DUNE as an example to measure the neutrino oscillation effects. In
    \cite{Kelly:2019itm}, studies has been performed to show CP sensitivity with sub-GeV atmospheric neutrinos at DUNE\cite{Kelly:2019itm}, but to the best of our knowledge, this is the first time that complete study of neutrino oscillation physics (CP, mass-ordering and octant of $\theta_{23}$) with the wide energy range (sub-GeV to 10 GeV) has been conducted for DUNE. 
    Considering the presently planned sequencing schedule for DUNE, where several Far Detector modules will be installed before the intense neutrino beam from FNAL becomes available, such measurements can be likely performed to a significant part already before the beam data reaches its full power.
\section{Neutrino oscillations with atmospheric neutrinos}
Within the three-generation paradigm, the neutrino mass and mixing is parameterised in terms of three mixing angles ($\theta_{12},\theta_{13}, \theta_{23}$), two mass-squared differences ($\Delta{m^{2}_{21}}$, $\Delta{m^{2}_{31}}$) and a complex CP violating phase ($\delta_{CP}$)\cite{Capozzi:2017ipn,Capozzi:2016rtj}. Currently, the unknowns in the standard oscillation sector are the mass ordering among the three neutrino mass states, the octant of the atmospheric mixing angle $\theta_{23}$, and the value of the CP violating phase $\delta_{CP}$.\\
When neutrinos travel through earth matter their coherent forward charged current scattering with the electrons of the earth matter leads to an extra effective contribution to the neutrino mass matrix. In atmospheric neutrino oscillations, generally the dominant mixing angle is $\theta_{23}$, and the relevant mass-squared difference is $\Delta{m^{2}_{31}}$. However, in the sub-GeV region, and for larger baselines, the neutrino evolution is dominated by $\Delta{m^{2}_{21}}$ term compared to the $\Delta{m^{2}_{31}}$ and $\Delta{m^{2}_{32}}$ term. To understand the impact of the CP phase on the oscillation probabilities, we first discuss the CP asymmetry in the transition probabilities.
\begin{figure}
		\centering
		\includegraphics[width=0.4\linewidth]{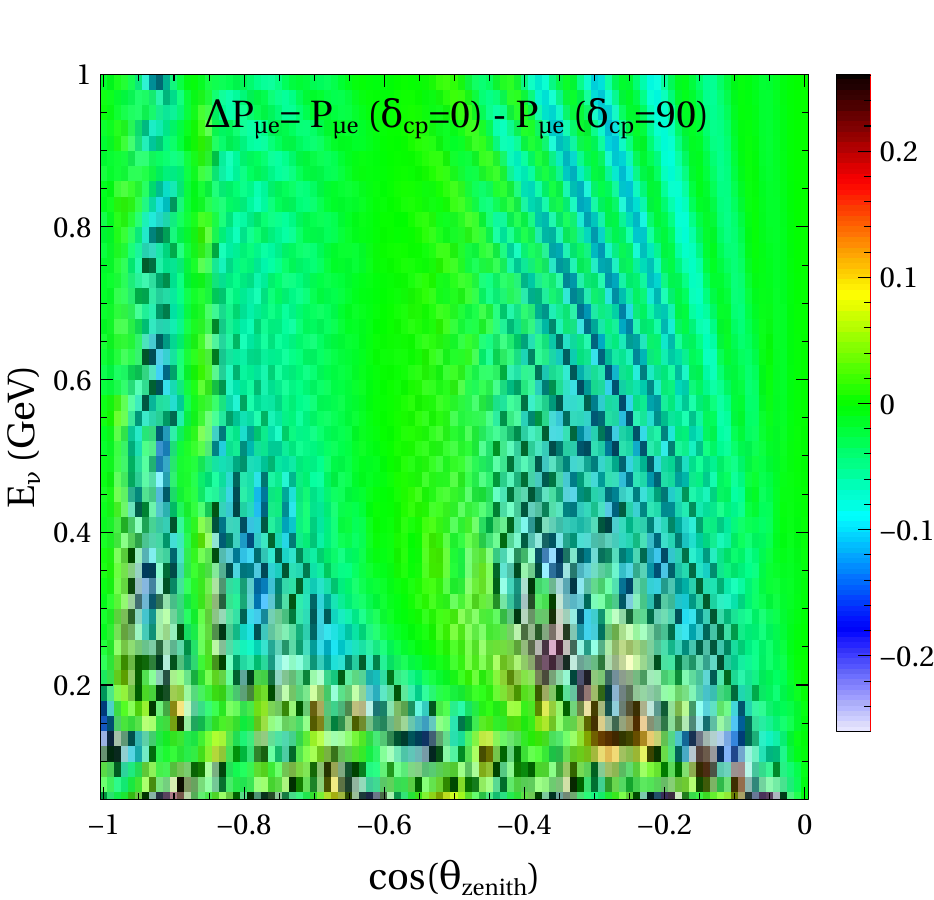}
		\includegraphics[width=0.4\linewidth]{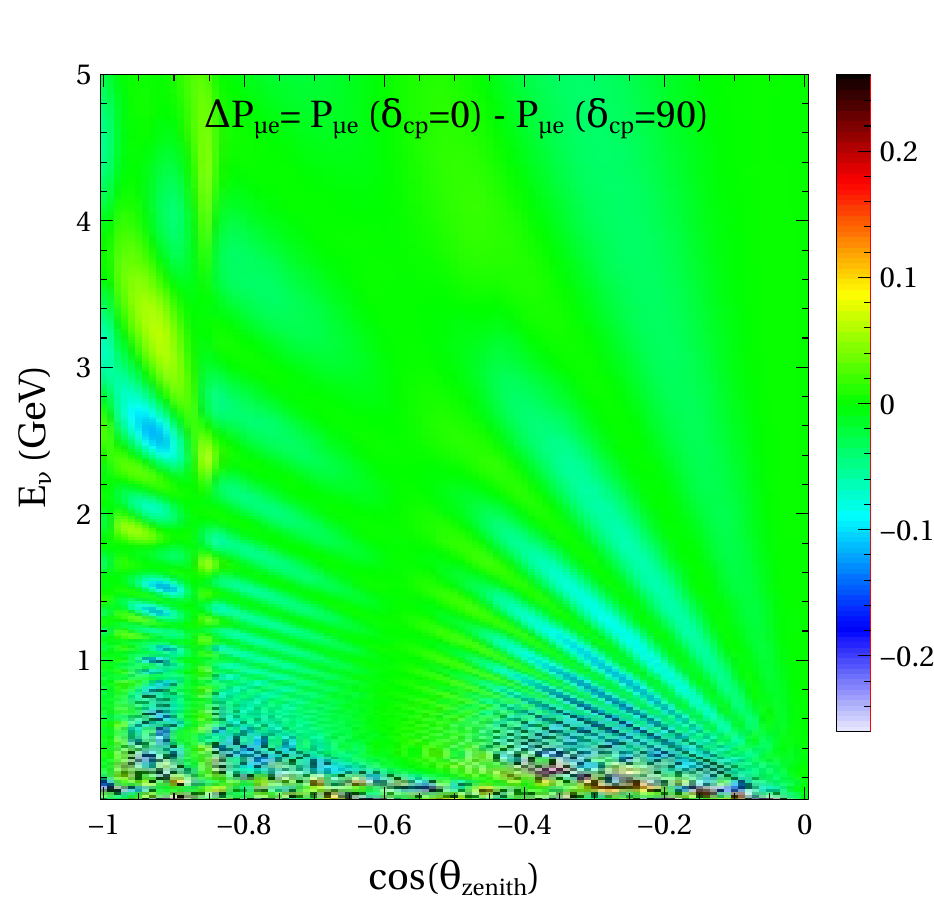}
		\includegraphics[width=0.4\linewidth]{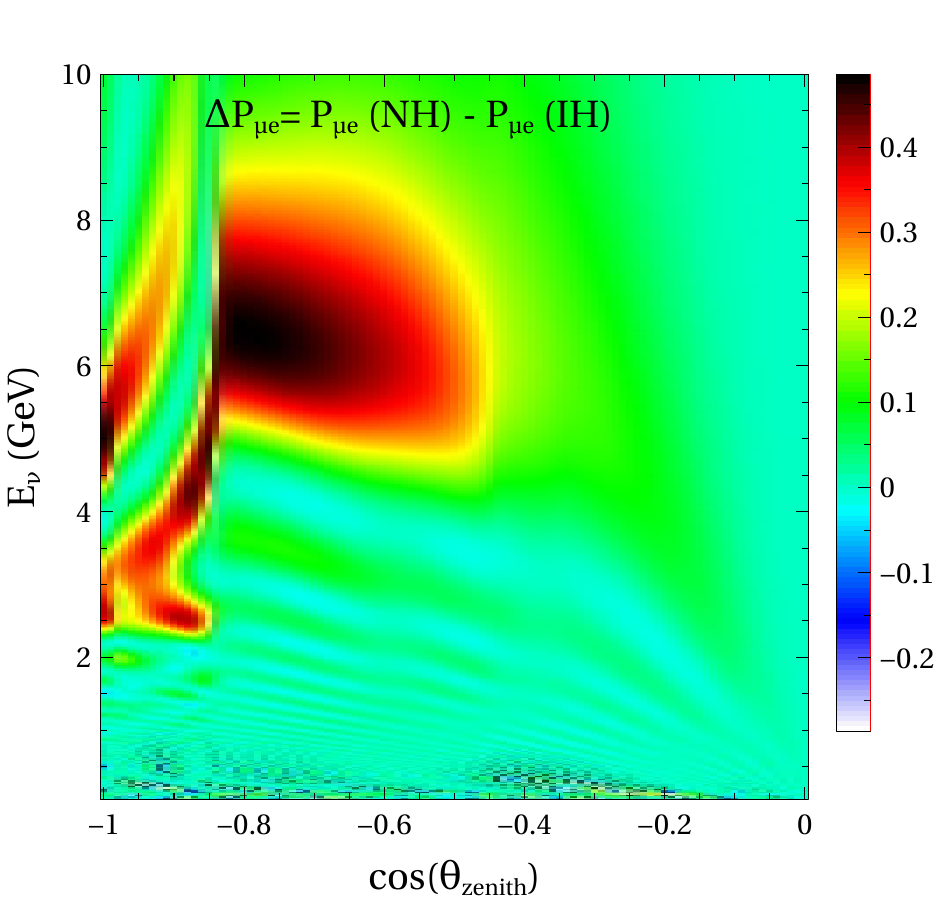}
		\includegraphics[width=0.4\linewidth]{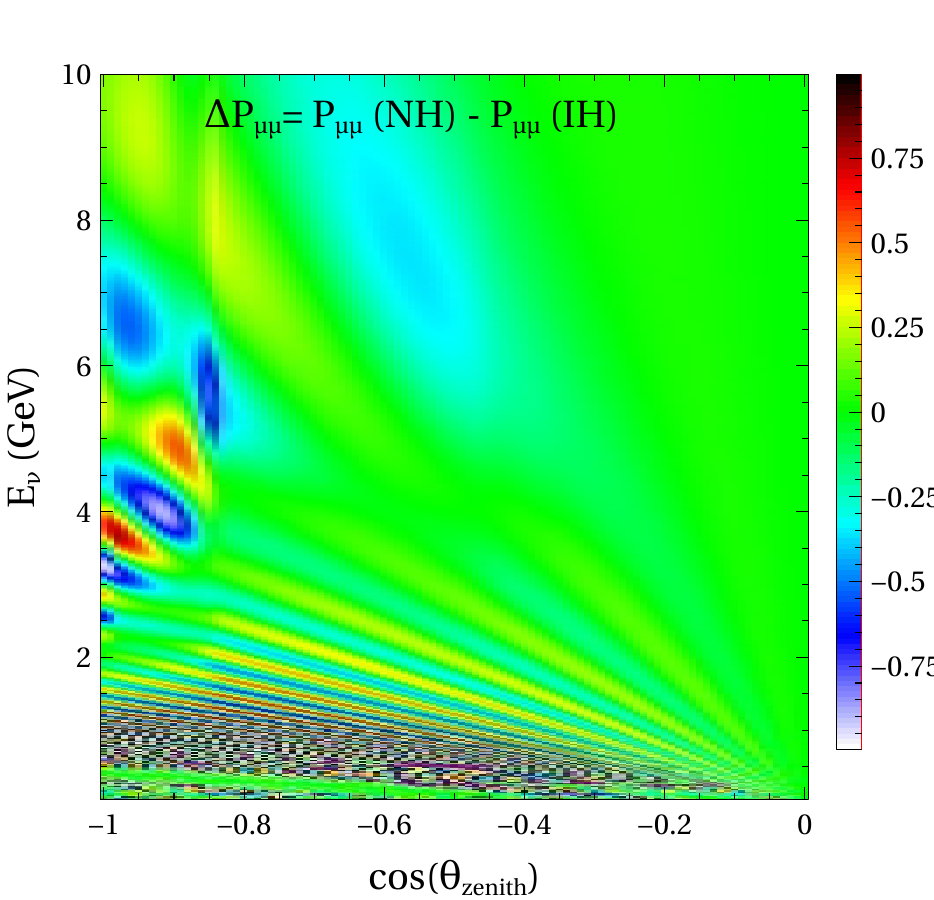}
        \includegraphics[width=0.4\linewidth]{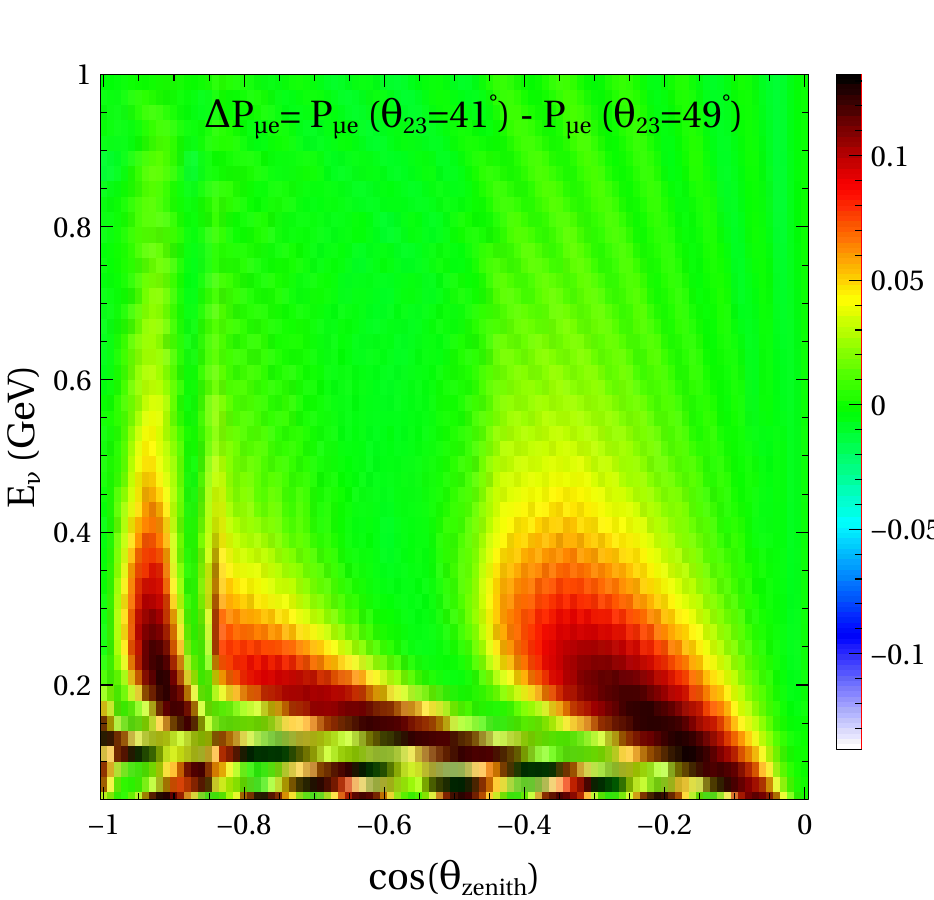}
		\includegraphics[width=0.4\linewidth]{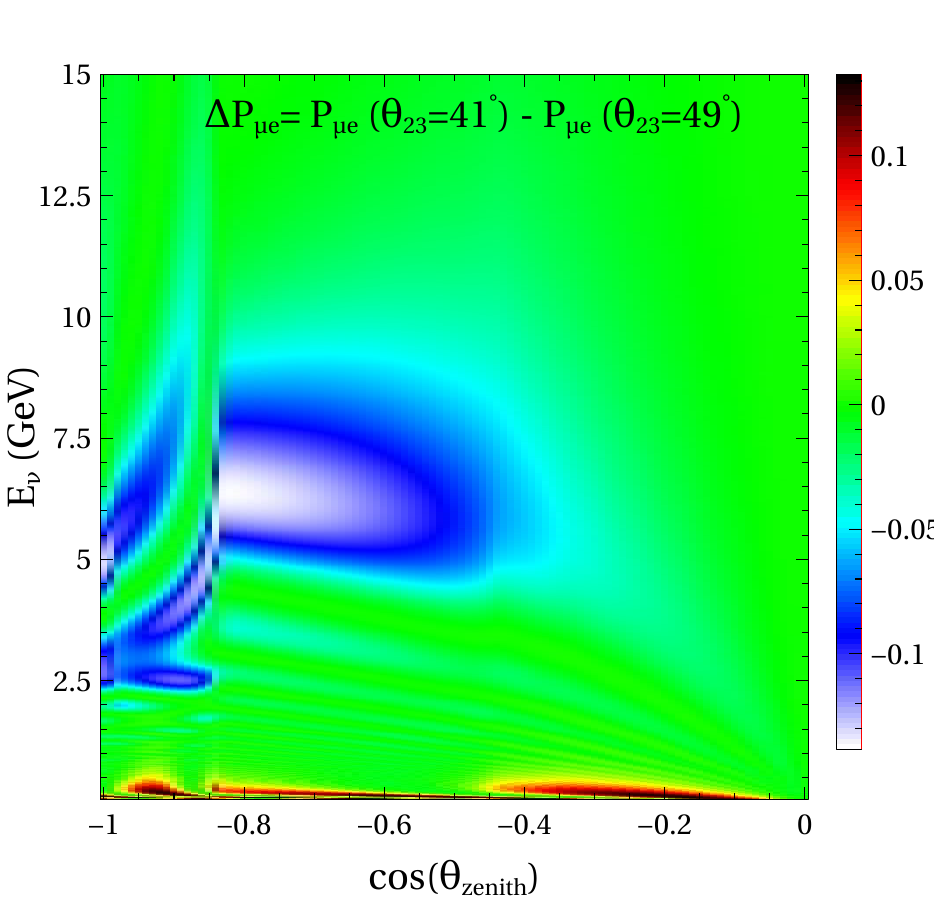}

		\caption{Top: Difference of the muon-appearance probability for a CP conserving ($\delta_{cp}=0$) and a CP violating ($\delta_{cp}=90^{o}$) phase is shown both for the sub-GeV (left) and high energy (right) region as a function of the neutrino energy and cosine of the zenith angle. Middle: Difference of the muon-appearance probability (left) and muon-disappearance probability (right) for the two different 
         mass-orderings is shown both as a function of neutrino energy and cosine of the zenith angle. Bottom: Difference of the muon-appearance probability for lower octant  ($\theta_{23}=41^{o}$) and higher octant ($\theta_{23}=49^{o}$) shown both for sub-GeV (left) and high energy(right) region as function of the neutrino energy and  cosine of the zenith angle.}
		\label{fig:pmm_terms}
	\end{figure}
	  The CP violating term in the transition probability can be written as \cite{Denton:2016wmg}
 \begin{equation}
     \Delta_{CP} = -8 {J_{r}} \sin(\delta_{cp}) \sin(\Delta_{12}) \sin(\Delta_{23}) \sin(\Delta_{13})
 \end{equation}
 where $\Delta_{ij}$ = $\delta{m^{2}_{ij}} L/4E$ and $J_{r}$=$c_{12} s_{12} c_{23} s_{23} c^{2}_{13} s^{2}_{13} $, with $c_{ij}=\cos\theta_{ij}$ and $s_{ij}=\sin\theta_{ij}$ respectively. To understand the impact of $\delta_{cp}$ on the transition probability as a function of energy, we need to separate the atmospheric neutrinos in two different energy region; i) $E_{\nu}$ is sub-GeV: In this energy range, the corresponding oscillatory terms average out whenever L/E is large compared to the $\Delta{m^{2}_{ij}}$. This leads to a much larger CP-violating contribution in the sub-GeV energy region. The top left panel of Fig.~\ref{fig:pmm_terms} shows the difference of the electron appearance channel probability for $\delta_{cp} =0$ and $\pi/2$ values. This clearly shows that the sub-GeV energy region will provide significant sensitivity to  CP violation effects from atmospheric neutrinos. ii) $E_{\nu}$ $>$ 1 GeV: in this energy range the oscillations are mostly dominated by the atmospheric mass splitting. Hence, the CP term will be suppressed by a factor of $\Delta{m^{2}_{21}}/ \Delta{m^{2}_{31}}$* $\pi/2$ $\approx$ 1/25 \cite{Peres:1999yi} as can be seen from 
 the top right plot of Fig.~\ref{fig:pmm_terms}.\\
    To describe the mass-ordering and octant sensitivity arising from atmospheric neutrinos, we use the appearance probability \cite{Elevant:2015ska,PhysRevD.64.053003}, which is valid for a constant matter density, and given by
\begin{equation}
        \begin{split}
           P_{{\mu}e}& \approx4\sin^2 \theta_{13}\sin^2 \theta_{23} \frac{\sin^2 \Delta_{31}(1-fmA)}{(1-fmA)^2}\\& + m*\frac{\Delta {m^2}_{21}}{\Delta {m^2}_{31}}\sin 2\theta_{13} \sin 2\theta_{12} \sin 2\theta_{23}\\
          &*\cos(m\Delta_{31} + f\delta_{cp}) \frac{\sin \Delta_{31}A}{A}\frac{\sin \Delta_{31} (1-fmA)}{(1-fmA)}  
        \end{split}
	\end{equation}
where $f=1$ for neutrinos and $f=-1$ for anti-neutrinos, $m$= sign($\Delta m^2 _{31}$), and $A=2EV/\Delta m^{2}_{31}$, where $V$ is the matter potential. It is seen that when atmospheric neutrinos travel inside the earth matter, they encounter sizeable changes due to the matter term which depends directly on the sign of $\Delta{m^{2}_{31}}$ and $\theta_{13}$. Hence, large matter effects with a wide energy range of atmospheric neutrinos will have a significant impact in resolving the mass-ordering. The difference of the two mass ordering schemes for $P_{{\mu}e}$ (left) and $P_{{\mu\mu}}$ (right) as a function of energy and cosine of the zenith angle is shown in the middle panel of Fig.~\ref{fig:pmm_terms}. The oscillogram plots clearly demonstrate that large matter effects play a crucial role in the appearance channel across the wide 
neutrino energy range. Hence, a mass-ordering sensitivity  arises both in the muon and electron channel from atmospheric neutrinos.\\
The octant of $\theta_{23}$ can mostly be resolved through the appearance probability, as the appearance channel depends on $\sin^2 \theta_{23}$. Hence, the measurement of appearance channel both for muon and electron will be ideal to resolve the degeneracy using the atmospheric neutrinos with a LArTPC detector such as  DUNE. The difference of muon-appearance probability for the lower octant  ($\theta_{23}=41^{o}$) and higher octant ($\theta_{23}=49^{o}$) is shown both for the sub-GeV (left) and high energy (right) region as a function of the cosine of zenith angle in the bottom panel of Fig.~\ref{fig:pmm_terms}. It is apparent that the sub-GeV atmospheric data will also help to resolve the octant degeneracy.
    
\section{Neutrino interactions and event topology within LArTPC detector}
Atmospheric neutrinos cover a broad range of energies in which neutrinos interact with the detector via charged or neutral current interactions. The atmospheric neutrino experiments measure the atmospheric flux at different energy scales, the importance of different interaction channels differ from one experiment to another. Hence, it is very important to evaluate the different interaction channels as they will have a significant impact on the measurement of oscillation parameters in each experiment in a unique way.\\
Here we have only considered  the charged-current neutrino interactions, which will be the dominant interaction channel for a neutrino experiment like DUNE. Charged-current neutrino interactions can be categorized in three major different types: a) Charged current quasielastic (CCQE) : CCQE interactions are mostly significant in the sub-GeV region ($<1 $GeV). In this interactions, the neutrino scatters off one of the bound nucleons and emits the charged-lepton partner. The outgoing nucleon is a proton for neutrinos, and a neutron for antineutrinos respectively. One can separate neutrinos and antineutrinos by identifying the proton in the final state. This 
method was also explored in \cite{Kelly:2019itm}. b) Resonance production (CCRES): the CCRES processes mostly dominate at slightly higher energies, up to 4 GeV. In this process, neutrinos (or anti-neutrinos) can excite an entire nucleon, and produce a $\Delta (1232)$, which subsequently decays into a pion nucleon pair. Similarly to CCQE, neutrinos (anti-neutrinos) produce more protons (neutrons) than neutrons (protons) in the final state. c) Deep inelastic scattering (DIS): DIS process mostly dominate above 4 GeV. In this process, neutrinos scatter off a single quark inside the nucleon and  produce the charged lepton plus a hadronic shower in the final state. In this type of interaction, it is difficult to separate the neutrino and antineutrino unless charged leptons are distinguished either by using a magnetic field or with statistical methods as described in\cite{Ternes:2019sak}, and separated from the hadronic shower. For this analysis
we have used only atmospheric neutrinos  with energies below 10 GeV, and explore the charge identification potential in the region of neutrinos with energies $< 4$ GeV.\\
So far, Super-Kamiokande and IceCube are the only large Cherenkov detectors, which are collecting atmospheric neutrinos with a broad energy spectrum. These experiments are not able to measure low-energy (sub-GeV) neutrinos precisely. It is extremely difficult to reconstruct the sub-GeV energies because of the poor reconstruction technique. For a Cherenkov detector, protons with less than 1.4 GeV do not emit any Cherenkov light in water, and hence can not be used to separate between neutrinos and antineutrinos.\\
The Liquid Argon Time Projection Chamber (LArTPC) with its full 3D-imaging, excellent particle identification (PID) capability and precise calorimetric energy reconstruction represents the most advanced experimental technology for neutrino detection for large detectors. Recently the ArgoNeut experiment \cite{Palamara:2016uqu, ArgoNeuT:2018tvi} has shown that protons with kinetic energy above 21 MeV can be reconstructed effectively. It has also shown that the detection of these protons allows the separation between sub-GeV neutrino and anti-neutrino interactions with Argon. Hence, together with the excellent energy resolution and identification of low energy proton, LArTPC 
detectors allow to separate neutrino (CC1P) and anti-neutrino (CC0P) events distinctively below 4 GeV at DUNE\cite{Andreopoulos:2015wxa,PhysRevD.102.112013,PhysRevLett.123.131801}.
\section{Experimental details}
    The atmospheric neutrino and anti-neutrino events are obtained by folding the relevant incident fluxes with the appropriate disappearance and appearance probabilities, charge current (CC) cross sections, detector efficiency, resolution, detector mass, and exposure time. 
        The $\mu^-$, and $e^{-}$ event rates in an energy bin of width $\mathrm{dE_\nu}$ and in a solid angle bin of width ${\mathrm{d \Omega_\nu}}$ are as follows:
    \begin{equation} \label{eq:muevent}
    \rm{ \frac{d^2 N_{\mu}}{d \Omega \;dE} = \frac{D_{eff}\Sigma}{2\pi} \left[\left(\frac{d^2 \Phi_\mu}{d \cos \theta \; dE}\right) P_{\mu\mu} +\left(\frac{d^2 \Phi_e}{d \cos \theta \; dE}\right)P_{e\mu}\right]}.
    \end{equation}
   \begin{equation} \label{eq:eevent}
    \mathrm{ \frac{d^2 N_e}{d \Omega \;dE} = \frac{D_{eff}\Sigma}{2\pi}\left[\left( \frac{d^2 \Phi_{\mu}}{d \cos \theta\; dE}\right)P_{\mu e} + \left(\frac{d^2 \Phi_e}{d \cos \theta \; dE}\right)P_{e e} \right]}
    \end{equation}

\begin{figure}
		\centering
		\includegraphics[width=0.48\linewidth]{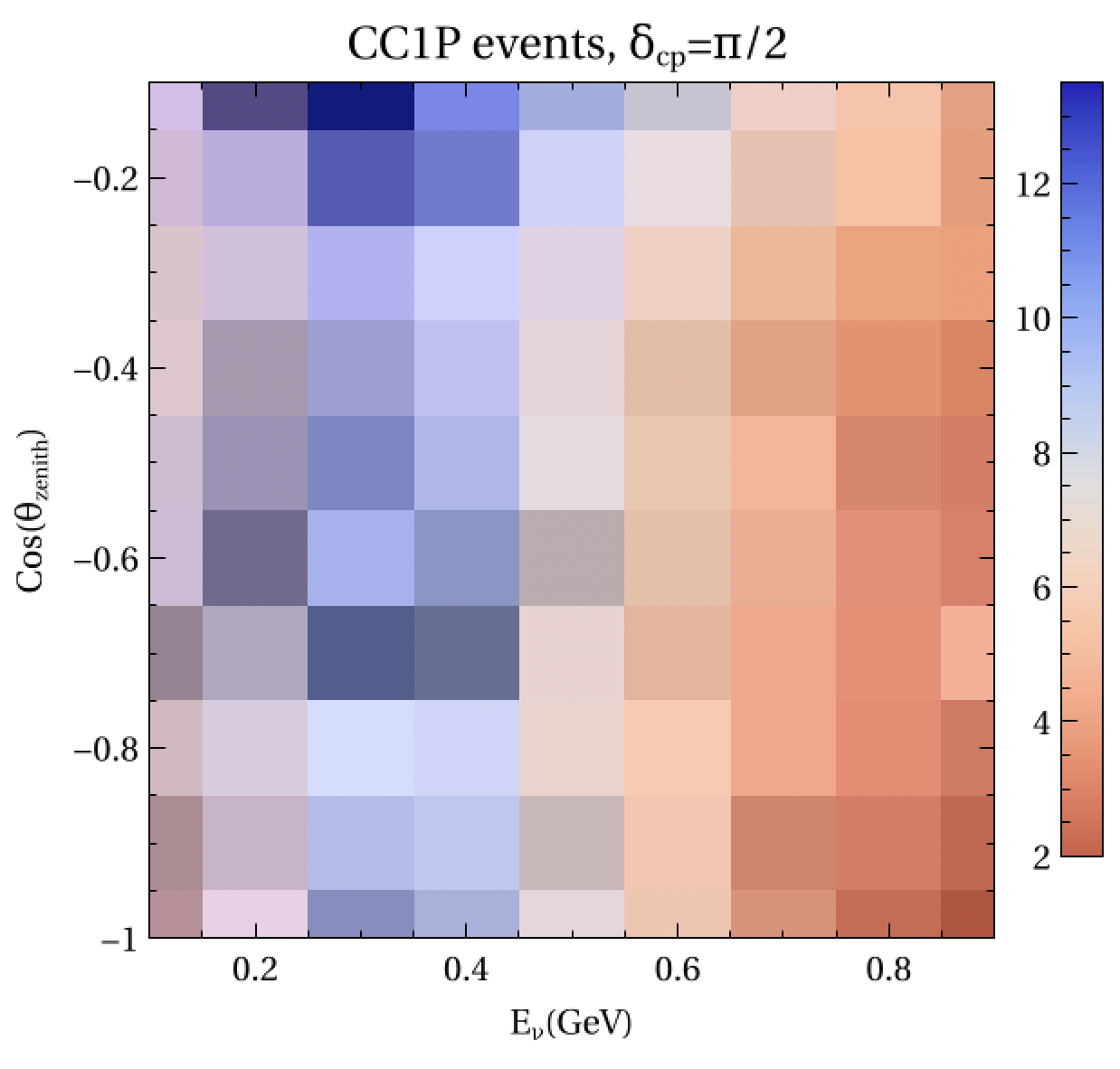}
		\includegraphics[width=0.48\linewidth]{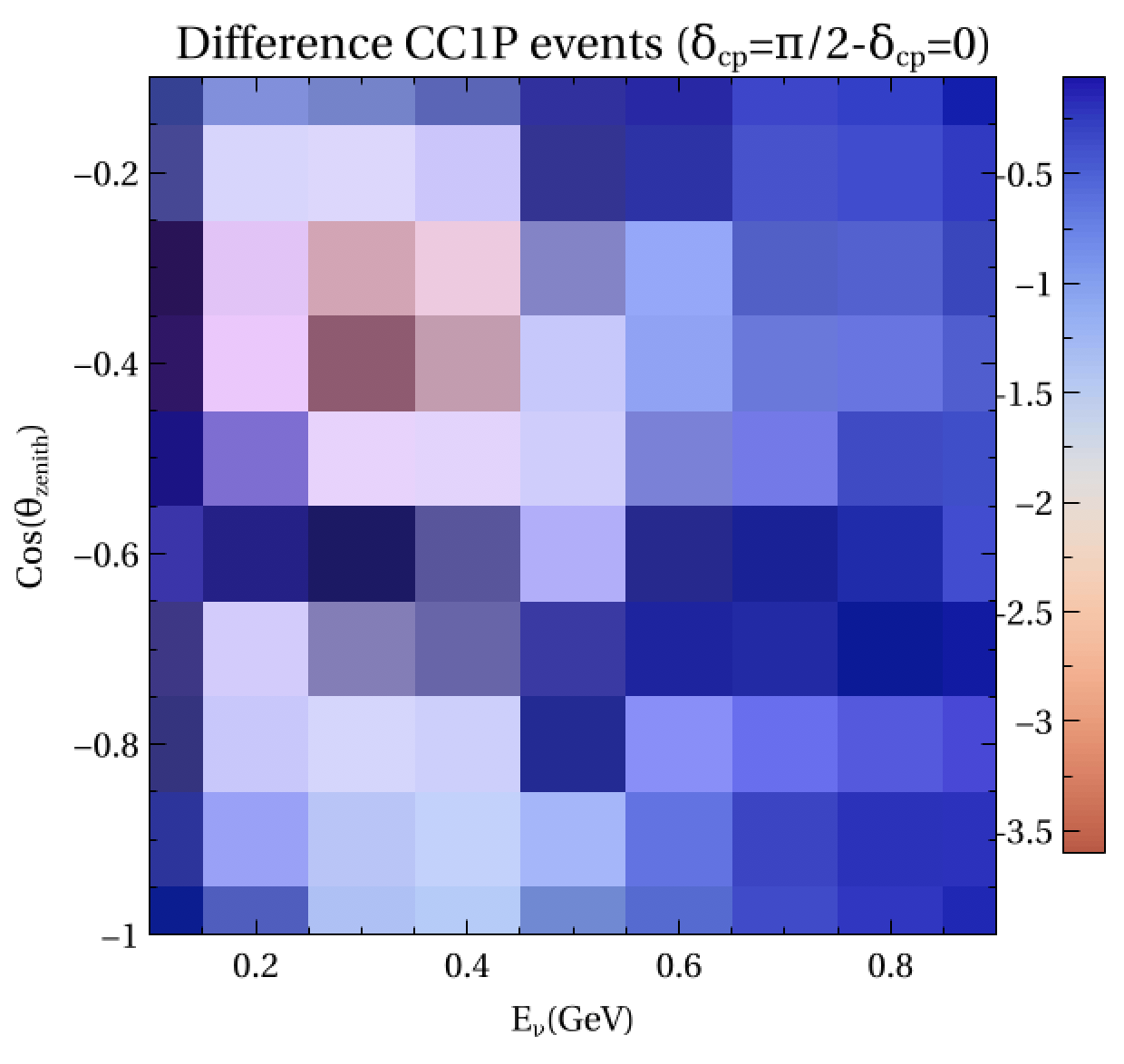}
		\caption{CC1P event distribution with true input value $\delta_{cp}=\pi/2$ (left) and the difference of CC1P events with $\delta_{cp}=0$ (right) as a function of neutrino energy (sub-GeV region) and $\cos(\theta_{zenith})$ for 140 kt-yr of atmospheric data only.}
		\label{fig:CC1Pevent}
	\end{figure}

    Here ${\mathrm{\Phi_{\mu}}}$ and ${\mathrm{\Phi_{e}}}$ are the $\mathrm{\nu_\mu}$ and ${\mathrm{\nu_e}}$ atmospheric fluxes  respectively, obtained from Honda et al.\cite{Hondaflux,Honda:2015fha} at the Homestake site; $P_{\mu\mu}(P_{ee})$ and $P_{\mu e}$ are disappearance and appearance probabilities; $\rm{\Sigma}$ is the total charge current (CC) cross-section taken from GENIE MC generator \cite{Andreopoulos:2009rq} and $\rm{D_{eff}}$ is the detector efficiency. The $\mu^+$, and $e^{+}$ event rates are similar to the above expression with the fluxes, probabilities, and cross sections replaced by those for $\mathrm{\bar\nu_\mu}$ and ${\mathrm{\bar\nu_e}}$ respectively.
    For a LArTPC detector, the energy and angular resolution are implemented using a Gaussian resolution function as follows,
    \begin{equation} \label{eq:esmear}
    \mathrm{ R_{E_\nu}(E_t,E_m) = \frac{1}{\sqrt{2\pi}\sigma} \exp\left[-\frac{(E_m - E_t)^2}{2 \sigma^2}\right]}\,.
    \end{equation}
    \begin{equation} \label{eq:anglesmear}
    \rm{ R_{\theta_\nu}(\Omega_t, \Omega_m) = N \exp \left[ - \frac{(\theta_t -\theta_m)^2 + \sin^2 \theta_t ~(\phi_t - \phi_m)^2}{2 (\Delta\theta)^2} \right] } \,,
    \end{equation}
    where N is a normalization constant. Here, $\rm{E_m}$ (${\rm{\Omega_m}}$), and $\rm{E_t}$ (${\rm{\Omega_t}}$) denote the measured and true values of energy (zenith angle) respectively. The smearing width $\sigma$ is a function of the energy $\rm{E_t}$. 
    Assumptions of the experimental smearing
    for the DUNE Far Detector (LArTPC) parameters for this study are reported in  \autoref{table:LAr-parameter} and taken from \cite{DUNE:2016ymp,DUNE:2018tke,Barr:2006it}.
        \begin{table}[H]
		\centering
		\begin{tabular}{|c|c|}
			\hline
			Parameter uncertainties or values & Value \\\hline
			$\mu^{+/-}$ angular uncertainty  & $3^\circ$\\
			$e^{+/-}$ angular uncertainty  & $5^\circ$\\
            proton angular uncertainty  & $10^\circ$\\
			$\mu^{+/-}$ energy uncertainty & 3$\%$ \\ 
           $e^{+/-}$ energy uncertainty & 5$\%$ \\
           proton energy uncertainty & 10$\%$ \\
           Detection efficiency  &  85$\%$\\
           Charge mis-identification efficiency  &  5$\%$\\
		      Flux normalization & 10$\%$\\ 
           Zenith angle uncertainty  & $25^\circ$\\
			 Cross section uncertainty & 20$\%$\\
			Additional overall systematic  & 5$\%$\\
			Flux Tilt\cite{Gonzalez-Garcia:2004pka}   & 5$\%$\\
		    \hline
		\end{tabular}
		\caption{Assumptions of the LArTPC Far Detector parameters and uncertainties.}
		\label{table:LAr-parameter}
	\end{table}
 Neutrino (anti-neutrino) events in the LArTPC detector are classified by event topology. We consider neutrino (anti-neutrino) events with a charged lepton (muon or electron) and one outgoing proton; for neutrinos the topology is used is CC1P and CC0P for anti-neutrinos. This technique is applied up to a neutrino energy of 4 GeV. The threshold kinetic energy for proton identification within LArTPC can be as low
 as 21 MeV as shown in \cite{Palamara:2016uqu}, but in our analysis we have used a more conservative energy threshold of 30 MeV\cite{DeRoeck:2020ntj,Kelly:2019itm}.  In this analysis the neutrino energy is reconstructed as $E_{\nu}$ = $E_{lepton}$ + $K_{proton}$, where $K_{proton}$ is the proton kinetic energy. Any neutrino event with proton energy higher than 30 MeV is included in this analysis. Neutrino and anti-neutrino events are summed (for the same flavour) for each energy and angular bins above 4 GeV. We  show the CC1P event distribution with true input value $\delta_{cp}=\pi/2$ (left) and the difference of CC1P events with $\delta_{cp}=0$ (right) as a function of neutrino energy (sub-GeV region) and $\cos(\theta_{zenith})$ for 140 kt-yr (see Table II) of atmospheric data in 
 Fig.~\ref{fig:CC1Pevent}. The right panel of Fig.~\ref{fig:CC1Pevent} shows that there are a total of 40 bins, which will provide significant sensitivity to the CP violation from the sub-GeV energy region.

     \section{ Analysis and results}
    The computation of $\chi^2$ is performed using the method of pulls. This method allows us to take into account the various statistical and systematic uncertainties in a straightforward way. The flux, cross sections and other systematic uncertainties are included by allowing these inputs to deviate from their standard values in the computation of the expected rate in the $\text{i-j}^{\text{th}}$ bin, ${\mathrm{ N^{th}_{ij} }}$. 
    
    \begin{equation}\label{eqn:cij}
    \mathrm{ N^{th}_{ij} =  N^{th}_{ij}(std) + \sum^{npull}_{k=1}
    \sigma_{ij}^k \xi_k }\,,
    \end{equation}
    
    where ${\mathrm{ N^{th}_{ij}(std) }}$ is the expected rate in the $\text{i-j}^{\text{th}}$ bin calculated with the standard values of the inputs.  $\sigma_{ij}^k$ and $\xi_k$ are the values of the uncertainties and the pull respectively. 
    The $\chi^2$ is calculated as described in \cite{Chatterjee:2014gxa}, which includes the effects of all theoretical and systematic uncertainties (as reported in \autoref{table:LAr-parameter}).
    In the case of the DUNE LArTPC detector, the $\chi^{2}$ with charge-id below 4 GeV neutrino energy is calculated as follows,
       \begin{equation}
           \chi^{2}_{< 4 {\rm GeV}} = \chi^{2}_{\mu^{-}}  + \chi^{2}_{ \mu^{+}} + \chi^{2}_{e^{-}} + \chi^{2}_{e^{+}}
    \end{equation}
    and $\chi^{2}$ without charge-id above 4 GeV neutrino energy 
    \begin{equation}
     \chi^{2}_{> 4 {\rm GeV}} = \chi^{2}_{\mu^{-} + \mu^{+}} + \chi^{2}_{e^{-}+ e^{+}}
    \end{equation}
 
    Finally, $\Delta{\chi^{2}}$ is marginalized over the oscillation parameters, given in Table III\cite{Gariazzo:2017fdh}. To calculate the experimental sensitivity, we have simulated events for up-going atmospheric neutrinos.
The assumptions on the collected data samples per year are given in Table II. Note that this is an optimistic scenario where, following the installation of the first Far Detector, each year another detector, up to four in total, is assumed to be added and leads to a data collection of
140 kt-year in five years. Presently, the Far Detectors 
three and four are likely to become ready somewhat later, so the first 5 years will collect perhaps more closer to 100 kt-year. But in any case the 
atmospheric neutrino program is expected to start several years earlier than the neutrino beam program.


\begin{table}[h]
    \centering
    \begin{tabular}{c c c}
        \begin{tabular}{|c|c|}
            \hline
             Year & kt-year \\
             \hline
        First  & 10\\
			Second  & 20\\
            Third  & 30\\
			Fourth & 40 \\ 
           Fifth & 40 \\
           Total & 140 \\
            \hline
        \end{tabular}
        &
           \begin{tabular}{|c|c|c|}
            \hline
             Parameters & True Values & Range \\
             \hline
        $\theta_{12}$ & $33.47^\circ$ & N.A.\\
			$\theta_{13}$ & $8.54^\circ$ & N.A. \\
            $\theta_{23}$ & $45^\circ$ & $41^\circ$: $49^\circ$\\
			$\Delta_{21}$ (eV$^2$) & $7.42\times 10^{-5}$  & N.A.\\ 
           $\Delta_{31}$ (eV$^2$) & $2.515\times 10^{-3}$  & $(2.41-2.61)10^{-3}$  \\
           $\delta_{cp}$ & 0 & $-180^\circ$: $180^\circ$ \\
            \hline
        \end{tabular}\\  
       Table II:Exposure& Table III: Oscillation parameters used
    \end{tabular}
\end{table}

\begin{figure}[H]
		\centering
		\includegraphics[width=0.58\linewidth]{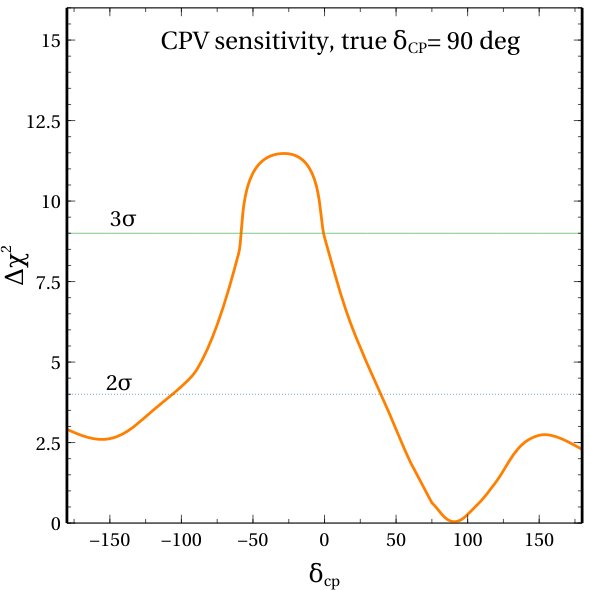}
		\caption{Sensitivity to the CP violating phase $\delta_{CP}$ using atmospheric neutrinos for true  $\delta_{CP}$ = 90$^o$ with an exposure of 140 kt-yr the DUNE LArTPC detector data.}
		\label{fig:CPV}
	\end{figure}

 \begin{figure}[H]
		\centering
		\includegraphics[width=0.48\linewidth]{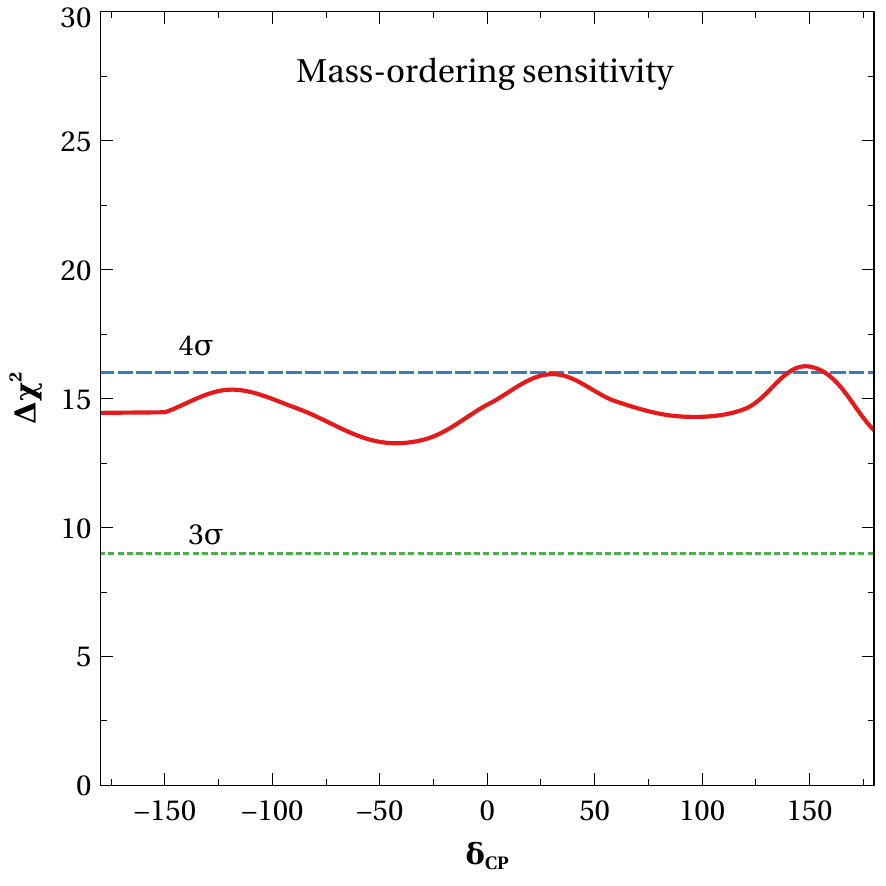}
		\includegraphics[width=0.48\linewidth]{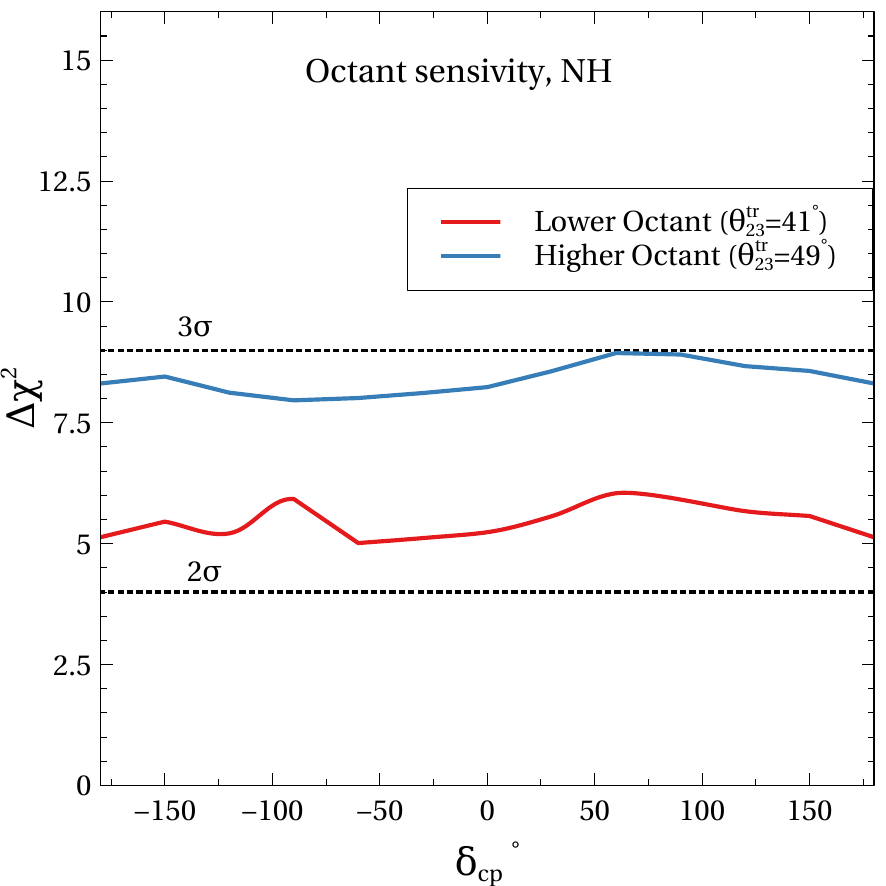}
		\caption{Sensitivity to the mass-ordering (left) and octant of $\theta_{23}$ (right) using atmospheric neutrinos with an exposure of 140 kt-yr of LArTPC detector data such as DUNE.}
		\label{fig:MH_Oct}
	\end{figure}
 The sensitivity to the CP violating phase $\delta_{CP}$ using only atmospheric neutrinos for a true  $\delta_{CP}$ = 90$^o$ with an exposure of 140 kt-yr of LArTPC data  at DUNE is shown in Fig.~\ref{fig:CPV}. The  result shows that a large region of $\delta_{CP}$ parameter space can be excluded with more than 2$\sigma$ and a fraction of $\delta_{CP}$ space with more than 3$\sigma$ with 140 kt-yr($\approx$ 5 years) of atmospheric data only. The sensitivity to the CPV phase can be understood as follows:  the CP violation effect is much larger for sub-GeV atmospheric neutrinos compared to the GeV beam neutrinos as explained in the previous section. The separation of an neutrino and anti-neutrino sample by using the event topology ($<$ 4 GeV) inside the LArTPC also improves the CP sensitivity. The combination of these two factors drive the CPV sensitivity at DUNE.\\
 In Fig.~\ref{fig:MH_Oct}, the sensitivity to the mass-ordering (left) and octant of $\theta_{23}$ (right) is shown using  atmospheric neutrinos with an exposure of 140 kt-yr for DUNE. A 
 significant sensitivity to the mass ordering may be achieved for all $\delta_{CP}$ values with a 
 sensitivity close to 4$\sigma$. The reason behind this very promising result is twofold. First, the large matter effect due to the long baseline enhances the sensitivity for all energies as shown in the middle panel of Fig.~\ref{fig:pmm_terms}. Second, the separation of neutrinos and anti-neutrinos within a LArTPC provides significant sensitivity to the mass-ordering within the energy range 1-4 GeV. Also notice that, in the case of normal (inverted) ordering, the matter resonance effects are at the aforementioned energies (2–8 GeV) for almost vertical upgoing ($-1 < \cos(\theta_{zenith}) < -0.5$) neutrinos (anti-neutrinos) respectively.  Hence, the combination of both large matter effect and charge separation helps to get a strong sensitivity to the mass-ordering.
 The sensitivity to the octant of $\theta_{23}$  for the higher (blue) and lower (red) octant is shown in right panel of Fig.~\ref{fig:MH_Oct}. The sensitivity of the higher and lower octant are  to a level of 3$\sigma$ and larger than 2$\sigma$, respectively. The octant sensitivity mostly arises from the appearance channel, $\nu_{\mu}\rightarrow \nu_{e}$ and $\nu_{e}\rightarrow \nu_{\mu}$ for atmospheric neutrinos across the wide energy range as can be seen from bottom panel of Fig.~\ref{fig:pmm_terms}.\\
 As a next step one could study the impact of using such atmospheric neutrino data in combination with 
 the data from the early neutrino beam, when these become 
 available a few years after the completion of the first Far Detector.
 In addition, one can try to  extend the 
 atmospheric neutrino analysis for the higher energy data ($>4$ GeV) including  statistical methods for the charge determination, which we 
 intend to study in detail in a forthcoming publication. However, it is clear that atmospheric neutrinos will provide a unique analysis of both CP violation, mass-ordering and octant determination of $\theta_{23}$ using the cutting-edge advanced technique of liquid argon time projection chambers and the wide energy range of atmospheric neutrinos in the initial data taking period of DUNE. 
\section{Conclusions}
Large LArTPCs are becoming available for the study of neutrino physics, and these detectors have excellent capabilities to reconstruct and classify neutrino interactions by topology, which give a sensitivity to neutrino and 
anti-neutrino interactions. This can be used to 
study neutrino oscillations and extract the CP violation, neutrino mass ordering and 
the octant determination of $\theta_{23}$ parameters. The results show that with a data 
sample collected over several years, one can 
achieve interesting sensitivities to these 
quantities, complementary to first neutrino beam results. These measurements can however start when the the first Far Detector of the DUNE experiment
is completed and starts collecting data.

\acknowledgments
    AC acknowledges CERN Neutrino Platform and the Ramanujan Fellowship (RJF/2021/000157), of the Science and Engineering Research Board of the Department of Science and Technology, Government of India. 
    ADR acknowledges CERN Neutrino platform.
    It is to be noted that this work has been done solely by the authors and is not representative of the DUNE collaboration.

\bibliographystyle{apsrev4-1}
\bibliography{references}

\end{document}